\documentclass[twocolumn,showpacs,prb]{revtex4}
\usepackage{graphicx}

\def\gz{\ifmmode{Z\hskip -4.8pt Z}
    \else{\hbox{$Z\hskip -4.8pt Z$}}\fi}

\newcommand{\be}{\begin{equation}}
\newcommand{\ee}{\end{equation}}
\newcommand{\bea}{\begin{eqnarray}}
\newcommand{\eea}{\end{eqnarray}}

\begin{document}

\title{Effect of covalency and interactions on the trigonal splitting in Na$_{x}$CoO$_{2}$}
\author{A.~A.~Aligia}
\email{aligia@cab.cnea.gov.ar}
\affiliation{Centro At\'{o}mico Bariloche and Instituto Balseiro, Comisi\'{o}n Nacional
de Energ\'{\i}a At\'{o}mica, 8400 Bariloche, Argentina}

\begin{abstract}
We calculate the effective trigonal crystal field $\Delta$ which splits the $t_{2g}$
levels of effective models for Na$_{x}$CoO$_{2}$ as the local symmetry
around a Co ion is reduced from $O_h$ to $D_{3d}$. To this end we solve
numerically a CoO$_6$ cluster containing a Co ion with all $3d$ states and
their interactions included, and its six nearest-neighbor O atoms, with the
geometry of the system, in which the CoO$_6$ octahedron is compressed along
a $C_3$ axis. We obtain $\Delta \approx 130$ meV, with the sign that agrees
with previous quantum chemistry calculations, but disagrees with
first-principles results in the local density approximation (LDA). We find
that $\Delta$ is very sensitive to a Coulomb parameter which controls the
Hund coupling and charge distribution among the $d$ orbitals. The origin of
the discrepancy with LDA results is discussed.
\end{abstract}

\maketitle

date{\today}


\section{Introduction}

The doped layered hexagonal cobaltates Na$_{x}$CoO$_{2}$ have attracted great
interest in the last years due to the high thermopower and at the same time
low thermal conductivity and resistivity for 0.5 $<x<$ 0.9, \cite{teras,mika}
and the discovery of superconductivity in hydrated Na$_{x}$CoO$_{2}$.\cite%
{taka} Further attention was motivated by the fact that first-principles
calculations in the local density approximation (LDA) \cite{lda,lda2,lda3}
predicted a Fermi surface with six prominent hole pockets along the $\Gamma
-K$ direction, which are absent in measured angle-resolved photoemission
(ARPES) spectra.\cite{arpes1,arpes2} To explain the discrepancy, several
calculations including correlation effects were made.\cite%
{zhou,ishi,anto,mari,lieb,gut,dmft} These studies used an effective model $%
H_{\mathrm{eff}}$ for the $t_{2g}$ $3d$ states of Co, split by the trigonal
crystal field $\Delta $ into an $a_{1g}^{\prime }$ singlet and an 
$e_{g}^{\prime }$ doublet.\cite{note} Except for some simplifications used in
the different works, $H_{\mathrm{eff}}$ has the form

\begin{eqnarray}
H_{\mathrm{eff}} &=&\sum_{i,\sigma }\Delta (\sum_{\beta \in e_{g}^{\prime }}%
\tilde{d}_{i\beta \sigma }^{\dagger }\tilde{d}_{i\beta \sigma }-\tilde{d}%
_{ia_{1g}^{\prime }\sigma }^{\dagger }\tilde{d}_{ia_{1g}^{\prime }\sigma }) 
\nonumber \\
&+&\sum_{i\delta \beta \gamma \sigma }\tilde{t}_{\delta }^{\beta \gamma }(%
\tilde{d}_{i+\delta ,\beta \sigma }^{\dagger }\tilde{d}_{i\gamma \sigma }+%
\mathrm{H.c.})+U_{\mathrm{eff}}\sum_{i\beta }\tilde{n}_{i\beta \uparrow }%
\tilde{n}_{i\beta \downarrow }  \nonumber \\
&+&\frac{1}{2}\sum_{i,\gamma \neq \beta ,\sigma \sigma ^{\prime }}(U_{%
\mathrm{eff}}^{\prime }\tilde{n}_{i\gamma \sigma }\tilde{n}_{i\beta \sigma
^{\prime }}+J_{\mathrm{eff}}\tilde{d}_{i\gamma \sigma }^{\dagger }\tilde{d}%
_{i\beta \sigma ^{\prime }}^{\dagger }\tilde{d}_{i\gamma \sigma ^{\prime }}%
\tilde{d}_{i\beta \sigma })  \nonumber \\
&+&J_{\mathrm{eff}}^{\prime }\sum_{\gamma \neq \beta }\tilde{d}_{i\gamma
\uparrow }^{\dagger }\tilde{d}_{i\gamma \downarrow }^{\dagger }\tilde{d}%
_{i\beta \downarrow }\tilde{d}_{i\beta \uparrow },  \label{heff}
\end{eqnarray}
where $\tilde{d}_{i\beta \sigma }^{\dagger }$ creates a hole on an 
\emph{effective} $t_{2g}$ orbital at site $i$ with spin $\sigma $. The first term
is the effective trigonal splitting mentioned above, the second term
describes the hopping between orbitals at a distance $\delta $ and the
remaining terms are effective interactions discussed for example in Ref. 
\onlinecite{effd}.

In most works, $\Delta$ and $\tilde{t}_{\delta }^{\beta \gamma }$ were derived 
from fits to the LDA bands and the interaction parameters were estimated. These
fits give either $\Delta =-10$ meV \cite{zhou} or $\Delta =-130$ meV.\cite{ishi} 
With these parameters and realistic values of the Coulomb repulsion $U_{\mathrm{eff}}$, 
correlations are not able to reconcile theory with
experiment, as shown by different dynamical-mean-field-theory (DMFT) 
studies.\cite{mari,lieb,dmft} The pockets still remain in the calculations.

Using instead an $H_{\mathrm{eff}}$ derived from a multiband Co-O model $H_{%
\mathrm{mb}}$ through a low-energy reduction procedure, \cite{effd} and the
value $\Delta =315$ meV obtained from quantum-chemistry
configuration-interaction calculations,\cite{ll} these pockets are absent
and the electronic dispersion near the Fermi energy agrees with experiment.%
\cite{dmft} In this procedure, no LDA results were used. The parameters of $%
H_{\mathrm{mb}}$ were taken from previous fits of \ of polarized x-ray
absorption spectra,\cite{kroll} and the parameters of $H_{\mathrm{eff}}$
other than $\Delta $ were obtained fitting the energy levels of an
undistorted CoO$_{6}$ cluster ($O_{h}$ symmetry) and calculating the
effective hopping between different CoO$_{6}$ clusters,\cite{effd} following
similar ideas that were successful in the superconducting cuprates.\cite%
{opti,erol,spife} In these systems, low-energy reduction procedures that
eliminate the O degrees of freedom, simplifying the problem to an effective
one-band one,\cite{opti,jef,schu,brin,bel,bel2,sys,hub} have been very
successful, in spite of the fact that doped holes enter mainly at O atoms 
\cite{nuck,kuip,pelle}. Optical properties related with O atoms were
calculated using these one-band models, which do not contain O states.\cite{opti,erol}

Summarizing previous results, if $\Delta$ is taken as a parameter, a positive 
$\Delta$ has the effect of shrinking the pockets, and for large enough $\Delta$,
the pockets disappear from the Fermi surface, reconciling 
theory with ARPES experiments.\cite{mari,lieb,dmft} 
A positive value has been obtained by quantum-chemistry
methods \cite{ll} and a negative one is obtained fitting the LDA dispersion with 
$H_{\mathrm{eff}}$.\cite{zhou,ishi} Thus, the origin of the discrepancy between different
methods and the actual value of $\Delta$ remains a subject of interest.

It is known that in general, the LDA underestimates gaps and has
difficulties in predicting one-particle excitations energies. Thus one might
suspect that the parameters of $H_{\mathrm{eff}}$, including $\Delta $
calculated with LDA are not accurate enough when covalency and interactions
are important. This is the case of NiO, for which agreement with experiment
in LDA+DMFT calculations is only achieved once the O bands are explicitly
included in the model,\cite{vol} or when the O atoms have been integrated
out using low-energy reduction procedures, which take into account
correlations from the beginning.\cite{vol,oles}

In covalent materials, the crystal-field splitting of transition-metal ions
is dominated by the hopping of electrons between these ions and their
nearest ligands.\cite{suga} In particular for Na$_{x}$CoO$_{2}$, an estimate
based on point charges gives $\Delta =-25$ meV.\cite{koshi} This shows that
the effect of interatomic repulsions is small and of the opposite sign as
that required to explain the ARPES spectra. The effects of covalency of Co
and its nearest-neighbor O atoms and all Co-Co interactions are included in
a CoO$_{6}$ cluster in the realistic ($D_{3d}$) symmetry. In this work, we
solve numerically this cluster and calculate the effective splitting $\Delta 
$, neglecting interatomic repulsions. We also analyze the
effects of different parameters on $\Delta $. The main result is that $%
\Delta \simeq 130$ meV and very sensitive to a parameter which controls the
Hund rules. It is also
sensitive to the cubic crystal-field splitting $10Dq$. A possible reason of
the discrepancy with the LDA results is discussed.

In Section II, we describe the model, parameters, and briefly the formalism.
Section III contains the results. Section IV is a summary and discussion.

\section{The model and its parameters}

The multiband model from which $H_{\mathrm{eff}}$ is derived, 
describes the $3d$ electrons of Co and the $2p$ electrons of the O atoms, 
located in the positions determined by the
structure of Na$_{0.61}$CoO$_{2}$ at 12 K.\cite{stru} 
In this work we restrict the calculation to a cluster of one 
Co atom and its six nearest-neighbor O atoms.
The relevant filling
for the calculation of $\Delta $ corresponds to formal valences Co$^{4+}$
and O$^{2-}$, or 41 electrons to occupy the $3d$ \ shell of Co and the $2p$
shells of the six O atoms. This corresponds to 5 holes in the CoO$_{6}$
cluster. Thus, it turns out to be simpler to work with hole operators (which
annihilate electrons) acting on the vacuum state in which the Co ion is in
the $3d^{10}$ configuration and the O ions are in the $p^{6}$ one. The most
important physical ingredients are the interactions inside the $3d$ \ shell 
$H_{I}$ and the Co-O hopping ($t_{j}^{\eta \xi }$ below), parameterized as
usual, in terms of the Slater-Koster parameters.\cite{slat}. We include a
cubic crystal field splitting $\epsilon _{t_{2g}}-\epsilon _{e_{g}}=10Dq$

The Hamiltonian for the Co$O_6$ cluster takes the form

\begin{eqnarray}
H_{\mathrm{mb}} &=&\sum_{\alpha \in e_{g},\sigma }\epsilon _{e_{g}}d_{\alpha
\sigma }^{\dagger }d_{\alpha \sigma }+\sum_{\beta \in t_{2g},\sigma
}\epsilon _{t_{2g}}d_{\beta \sigma }^{\dagger }d_{\beta \sigma }+H_{I} 
\nonumber \\
&&+\sum_{j\eta \sigma }\epsilon _{\text{O}}p_{j\eta \sigma }^{\dagger
}p_{j\eta \sigma }+\sum_{j\eta \xi \sigma }t_{j}^{\eta \xi }(p_{j\eta \sigma
}^{\dagger }d_{\xi \sigma }+\mathrm{H.c.})  \label{hmb}
\end{eqnarray}%
The operator $d_{\xi \sigma }^{\dagger }$ creates a hole on the orbital $\xi 
$ of Co with spin $\sigma $. Similarly $p_{j\eta \sigma }^{\dagger }$
creates a hole on O $2p$ orbital $\eta $ at site $j$ with spin $\sigma $.
The first two terms corresponds to the energy of the $e_{g}$ orbitals ($%
x^{2}-y^{2}$, $3z^{2}-r^{2}$) and $t_{2g}$ orbitals ($xy$, $yz$, $zx$)
written on a basis in which $x$, $y$, $z$, point to the vertices of a
regular CoO$_{6}$ octahedron (symmetry $O_{h}$). The compression along the
axis $x+y+z$ reduces the symmetry to $D_{3d}$ and splits the states of
symmetry $xy+yz+zx$ ($a_{1g}^\prime$ in $D_{3d}$ \cite{note}) from the other
two $t_{2g}$ ones ($e_g^\prime$ in $D_{3d}$).

$H_{I}$ contains all interactions between $d$ holes assuming spherical
symmetry [the symmetry is reduced to $O_{h}$ by the cubic crystal field 
$10Dq $ and to $D_{3d}$ by the last (hopping) term of Eq. (\ref{hmb})]. The
expression of $H_{I}$ is lengthy. It is included in the Appendix [Eq. (\ref{hi})]
together with a brief description of its derivation for the interested reader.
A more detailed discussion is in Ref. \onlinecite{effd}.
The form of $H_{I}$  is rather simple and well
known when either only $e_{g}$ orbitals \cite{epl} or only $t_{2g}$ orbitals [as in Eq. (\ref{heff})] 
\cite{fre,gus} are important, although the correct expressions were not
always used.\cite{com,fre} In the general case, $H_{I}$ contains new terms which are
often disregarded. For example in a recent study of Fe pnictides,
\cite{kane} a simplified expression derived previously \cite{oles2} was used.
More recently, to estimate the effective Coulomb interaction for
transition-metal atoms on metallic surfaces, only density-density
interactions were included.\cite{gard} Some of the effects of these
simplifications were discussed in Ref. \onlinecite{effd}.

All interactions are given in terms of three free parameters $F_{0} \gg F_2 \gg F_{4}$. 
For example the Coulomb repulsion between two holes or
electrons at the same $3d$ orbital is $U=F_{0}+4F_{2}+36F_{4}$, and the Hund
rules exchange interaction between two $e_{g}$ ($t_{2g}$) electrons is 
$J_{e}=4F_{2}+15F_{4}$ ($J_{t}=3F_{2}+20F_{4}$). Thus $F_2$ is the main parameter responsible 
for the spin and orbital polarizations related with the first and second
Hund rules respectively.

Note that in Eq. (\ref{hmb}) there is no trigonal splitting. This means that
we take the \emph{bare} value of the splitting $\Delta_0=0$ (neglecting the effect 
of interatomic repulsions). The \emph{dressed} value $\Delta$ that enters
the effective Hamiltonian Eq. (\ref{hmb}) is calculated as

\be
\Delta=E(e_{g}^{\prime })-E(a_{1g}^{\prime }),
\label{del}
\ee
where $E(\Gamma)$ is the energy of the lowest lying state that transforms 
under symmetry operations according to the irreducible representation 
$\Gamma$ of the point group $D_{3d}$.\cite{note}

As in previous calculations for the regular CoO$_{6}$ octahedron (with
symmetry $O_{h}$),\cite{effd} the diagonalization is simplified by the fact
that several linear combinations of O $2p$ orbitals do not hybridize with
the Co $3d$ ones, forming non-bonding orbitals. However in the present case,
the reduced $D_{3d}$ symmetry increases the bonding $2p$ combinations to
seven, and a different basis should be used, but still the size of the
relevant Hilbert space is small enough to permit the diagonalization
numerically by the Lanczos method.\cite{gag}

As a basis for the present study, we take parameters determined previously 
\cite{kroll} from a fit of polarized x-ray absorption spectra of Na$_{x}$CoO$%
_{2}$ to the results of a CoO$_{6}$ cluster with 4 and 5 and holes including
the core hole. In the present case, we have neglected the O-O hopping for
simplicity (this allows a reduction of the relevant Hilbert space). Thus,
the parameters of $H_{\mathrm{mb}}$ in eV are \cite{kroll}

\begin{eqnarray}
F_{0} &=&3.5\text{, }F_{2}=0.2\text{, }F_{4}=0.006\text{, }  \nonumber \\
\epsilon _{\text{O}} &=&13\text{, }\epsilon _{t_{2g}}=1.2,\epsilon
_{e_{g}}=0,  \nonumber \\
(pd\pi ) &=&\frac{-\sqrt{3}}{4}(pd\sigma )=1.  \label{par}
\end{eqnarray}
The choice of the origin of on-site energies $\epsilon _{e_{g}}=0$ is
arbitrary. The resulting values of $U=4.516$ eV and charge transfer energies
are similar to those derived from other x-ray absorption experiments.\cite%
{wu} We note that while above $\epsilon _{t_{2g}}-\epsilon _{e_{g}}=10Dq=1.2$
eV, the effect of hybridization increases the splitting between $t_{2g}$ and 
$e_{g}$ orbitals to more than 3 eV.

\section{Results}

The splitting $\Delta$ is determined from Eq. (\ref{del}). We have also calculated the occupation of 
the $a_{1g}^\prime$ $3d$ orbital in each state to verify that the
expected physics is obtained.

For the parameters determined previously [Eq. (\ref{par})], we obtain 
$\Delta=124$ meV. The sign agrees with quantum-chemistry
configuration-interaction calculations \cite{ll} which obtained $\Delta
\approx 300$ meV, although our magnitude is smaller. The difference might be
at least partially due to some uncertainty in our parameters determined from
a fitting procedure. Motivated by this possibility, we have studied the effect of different
parameters on the results. Of course, since we have neglected interatomic
interactions, $\Delta$ vanishes if the hopping parameters $pd\sigma$ and 
$pd\pi$ are zero, and one would expect than an increase in these parameters, has
the largest impact on $\Delta$. However, we find that an increase of 50\% in
the hopping increases $\Delta$ by only 25\%. In addition, changes of the
oxygen energy $\epsilon _{\text{O}}$ (the charge transfer energy) or $F_0$
(which determines the intra-orbital Coulomb repulsion $U$) by 1 eV have an
effect of only a few percent on $\Delta$.

\begin{figure}[!ht]
\includegraphics[width=\linewidth]{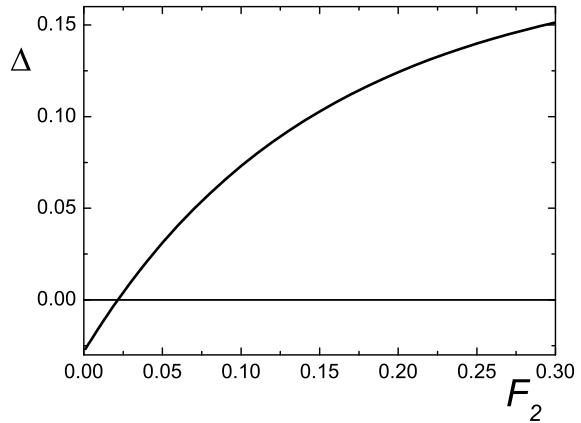} 
\caption{Trigonal splitting as a function of $F_2$ keeping the remaining
parameters as given by Eq. (\protect\ref{par}).}
\label{f2}
\end{figure}

Instead, and rather surprisingly, as shown in Fig. \ref{f2}, $\Delta$ is
very sensitive to $F_2$, the most important parameter in the expressions
for the exchange between $d$ electrons [$J_\nu$ with $\nu=e$, $t$, $a$ or $b$
in Eq.  (\ref{hi})] and the inter-orbital repulsions ($U-2J_\nu$) among other
interactions. Thus, it is the main responsible for the spin and orbital polarizations 
resulting in the first and second Hund rules. In particular, the repulsion between
different $e_g$ ($t_{2g}$) orbitals is reduced with respect to the
intra-orbital repulsion $U$ by $2J_e$ ($2J_t$) (see the Appendix). 

$\Delta$ becomes negative for $F_2 < 21$
meV. Curiously, increasing $F_4$ has a small effect, but in the 
\emph{opposite} sense as increasing $F_2$. This points to non-trivial effects of
the correlations, particularly those involving both $e_g$ and $t_{2g}$
electrons. When both $F_2$ and $F_4$ vanish we obtain a small
positive value $\Delta=12$ meV. If one adds to this result the contribution
-25 meV from the interatomic Coulomb repulsion estimated using point charges,\cite{koshi} 
one obtains a value close to -10 meV, obtained in one of the
LDA calculations.\cite{zhou} This suggest that the LDA negative results for 
$\Delta$ might be due to the difficulties of LDA in treating correlations
related with the Hund rules. In particular, it is known that orbital-related
Coulomb interactions are underestimated in the spin LDA,\cite{esch} and
empirical orbital polarization corrections \cite{erik} are frequently used
to cure this problem. This fact has been also analyzed in the framework
of a self-consistent tight-binding theory \cite{nico}

\begin{figure}[!ht]
\includegraphics[width=\linewidth]{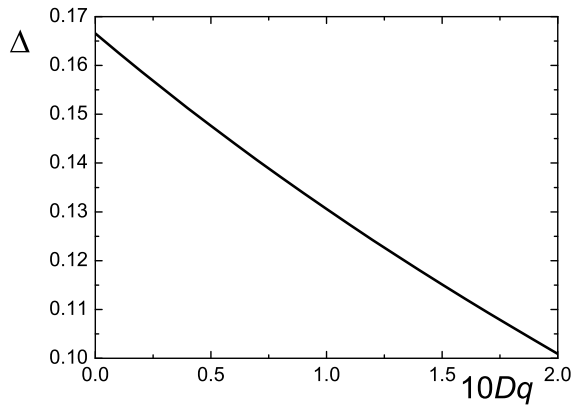}
\caption{Trigonal splitting as a function of the cubic crystal field keeping
the remaining parameters as given by Eq. (\protect\ref{par}).}
\label{ccf}
\end{figure}

The fact that correlations between both $e_g$ and $t_{2g}$ holes play a role
is supported by the dependence of $\Delta$ on the cubic crystal field
parameter $10Dq$, displayed in in Fig. \ref{ccf}.
Note that this parameter in the present case represents only
the contribution of interatomic repulsion to $10Dq$. The
covalency part is included in our calculation and the splitting between
hybridized $e_g$ and $t_{2g}$ is larger than 3 eV. Also in the fitting
procedure, the best value of $10Dq$ depends on composition $x$, being 1.2 eV
for $x=0.4$ and 0.9 eV for $x=0.6$.\cite{kroll} For the latter value $\Delta$
increases to 134 meV. As it is apparent in Fig. \ref{ccf}, $\Delta$
increases with decreasing $10Dq$. This shows that the $e_g$ states play an
important role. In fact, the results for the regular octahedron show that
although these states are absent in the effective Hamiltonian for the
cobaltates, they have a larger degree of covalency than the $t_{2g}$ states.%
\cite{effd} Most of the O holes reside in bonding combinations of $e_g$
symmetry.

\section{Summary and discussion}

Using exact numerical diagonalization of a CoO$_6$ cluster, with the 
realistic geometry of Na$_{x}$CoO$_{2}$, we have
calculated the effects of covalency and interactions on the trigonal
crystal-field parameter $\Delta$, which splits the $t_{2g}$ states in $O_h$
symmetry into $a_{1g}^\prime$ and $e_g^\prime$ in the reduced $D_{3d}$
symmetry. This parameter enters effective models [of the form of Eq. (\ref{heff})] 
for the description of the
electronic structure of Na$_{x}$CoO$_{2}$ and only positive values (in
contrast to the negative ones obtained from LDA) seem consistent with ARPES
data.\cite{mari,lieb,dmft}. We obtain $\Delta \approx 130$ meV.

While changes of the order of 1 eV in charge-transfer energy or $F_0$ 
(which controls the part of the Coulomb
repulsion which does not depend of the symmetry of the orbitals) 
do not affect $\Delta$ very much, we find that $\Delta$ is
very sensitive to the parameter $F_2$ which controls (among others) 
interaction constants related with the Hund rules (exchange interactions and
decrease of inter-orbital repulsions with respect to intra-orbital ones). To
a smaller extent, it is also sensitive to the cubic crystal field $10Dq$
reflecting the importance of interactions between $t_{2g}$ and $e_{g}$
states, and the effect on the latter on the effective parameters.

Since the LDA underestimates correlations that affect the orbital
polarization of the $d$ states,\cite{esch,erik,nico} this is likely to be the reason
of the failure of LDA approaches and effective models based on LDA
parameters, to reproduce the observed ARPES data. In fact, since 
the exchange and correlations in LDA are based on a homogeneous electron gas, 
it is expected that this approximation treats $F_0$ (the part of the repulsion
which does not distinguish between different orbitals) in mean field, but 
does not contain the effects of $F_2$ and $F_4$, which depend on the particular orbitals.
The exchange of the electron gas taken into account in the LDA helps to
follow the first Hund rule (maximum spin), but the second one, related
with orbital polarization, is not well described and seems crucial to establish
effective energy differences between different orbitals inside an incomplete
$d$ shell.

\section*{Acknowledgments}

Useful comments of G. Pastor and C. Proetto are thankfully acknowledged. AAA is partially
supported by CONICET, Argentina. This work was sponsored by PIP
112-200801-01821 of CONICET, and PICT 2010-1060 of the ANPCyT.

\appendix

\section{Interactions inside a $d$ shell}

\label{appe}

The part of the Hamiltonian that contains the interaction among the 10 d
spin-orbitals is \cite{nege}

\begin{equation}
H_{I}=\frac{1}{2}\sum_{\lambda \mu \nu \rho }V_{\lambda \mu \nu \rho
}d_{\lambda }^{+}d_{\mu }^{+}d_{\rho }d_{\nu },  \label{hi0}
\end{equation}%
where $d_{\lambda }^{+}$ creates an electron or a hole at \ the spin-orbital 
$\lambda $ ($H_{I}$ is invariant under an electron-hole transformation) and
(neglecting screening by other electrons)

\begin{equation}
V_{\lambda \mu \nu \rho }=\int d\mathbf{r}_{1}d\mathbf{r}_{2}\bar{\varphi}%
_{\lambda }(\mathbf{r}_{1})\bar{\varphi}_{\mu }(\mathbf{r}_{2})\frac{e^{2}}
{|\mathbf{r}_{1}-\mathbf{r}_{2}|}\varphi _{\nu }(\mathbf{r}_{1})\varphi _{\rho
}(\mathbf{r}_{2}),  \label{integ}
\end{equation}%
where $\varphi _{\lambda }(\mathbf{r}_{1})$ is the wave function of the
spin-orbital $\lambda $. Assuming spherical symmetry, these integrals can be
calculated using standard methods of atomic physics \cite{cond} in terms of
three independent parameters $F_{j}$, $j=0,2,4$, which are related to decomposition
of the the Coulomb interaction $e^2/|\mathbf{r}_{1}-\mathbf{r}_{2}|$ 
in spherical harmonics of degree $j$. To remove uncomfortable denominators, 
the three free parameters are defined as $F_{0}=R^{0}$, $F_{2}=R^{2}/49$ and $F_{4}=R^{4}/441$, 
where 
\be
R^{k} = e^{2}\int_{0}^{\infty }\int_{0}^{\infty }\frac{r_{<}^{k}}{%
r_{>}^{k+1}}R^{2}(r_{1})R^{2}(r_{2})r_{1}^{2}r_{2}^{2}dr_{1}dr_{2},
\label{r}
\ee
$R(r)$ is the radial part of the wave funcion of the $d$ orbitals and 
$r_{<}$ ($r_{>}$) is the smaller (larger) between $r_{1}$ and $r_{2}$.
The angular integrals are given in terms of tabulated coefficients.\cite{effd,cond}
Screening reduces $F_0$ significantly, but not $F_2$ and $F_4$.

The final result can be written
in the form below.\cite{effd} To express it in a more compact form, we
introduce different sums which run over a limited set of orbitals as
follows. The sums over $\alpha $  run over the five $d$ orbitals, those over 
$\beta ,\gamma $ run only over the $t_{2g}$ orbitals $xy$, $yz$, $zx$, and
those over $\chi $ ($\zeta $) run over the pair of orbitals $x^{2}-y^{2}$, $xy$ ($zx,zy$). 

The values of the different interactions energies below are given in terms of the
$F_j$ as follows: $U=F_{0}+4F_{2}+36F_{4}$, $J_{e}=4F_{2}+15F_{4}$, $%
J_{t}=3F_{2}+20F_{4} $, $J_{a}=35F_{4}$, $J_{b}=F_{2}+30F_{4}$, and $\lambda
=\sqrt{3}(F_{2}-5F_{4})$.

The interaction is

\begin{eqnarray}
&&H_{I}=U\sum_{\alpha }n_{\alpha ,\uparrow }n_{\alpha ,\downarrow } 
\nonumber \\
&+&(U-2J_{e})\sum_{\chi }\sum_{\sigma _{1},\sigma _{2}}n_{\chi ,\sigma
_{1}}n_{3z^{2}-r^{2},\sigma _{2}}  \nonumber \\
&+&\frac{U-2J_{t}}{2}\sum_{\beta \neq \gamma }\sum_{\sigma _{1},\sigma
_{2}}n_{\beta ,\sigma _{1}}n_{\gamma ,\sigma _{2}}  \nonumber \\
&+&(U-2J_{t})\sum_{\zeta }\sum_{\sigma _{1},\sigma
_{2}}n_{x^{2}-y^{2},\sigma _{1}}n_{\zeta ,\sigma _{2}}  \nonumber \\
&+&(U-2J_{a})\sum_{\sigma _{1},\sigma _{2}}n_{x^{2}-y^{2},\sigma
_{1}}n_{xy,\sigma _{2}}  \nonumber \\
&+&(U-2J_{b})\sum_{\zeta }\sum_{\sigma _{1},\sigma
_{2}}n_{3z^{2}-r^{2},\sigma _{1}}n_{\zeta ,\sigma _{2}}  \nonumber \\
&+&J_{e}\sum_{\chi }\sum_{\sigma _{1},\sigma _{2}}d_{\chi ,\sigma
_{1}}^{\dagger }d_{3z^{2}-r^{2},\sigma _{2}}^{\dagger }d_{\chi ,\sigma
_{2}}d_{3z^{2}-r^{2},\sigma _{1}}  \nonumber \\
&+&\frac{J_{t}}{2}\sum_{\beta \neq \gamma }\sum_{\sigma _{1},\sigma
_{2}}d_{\beta ,\sigma _{1}}^{\dagger }d_{\gamma ,\sigma _{2}}^{\dagger
}d_{\beta ,\sigma _{2}}d_{\gamma ,\sigma _{1}}  \nonumber \\
&+&J_{t}\sum_{\zeta }\sum_{\sigma _{1},\sigma _{2}}d_{x^{2}-y^{2},\sigma
_{1}}^{\dagger }d_{\zeta ,\sigma _{2}}^{\dagger }d_{x^{2}-y^{2},\sigma
_{2}}d_{\zeta ,\sigma _{1}}  \nonumber \\
&+&J_{a}\sum_{\sigma _{1},\sigma _{2}}d_{x^{2}-y^{2},\sigma _{1}}^{\dagger
}d_{xy,\sigma _{2}}^{\dagger }d_{x^{2}-y^{2},\sigma _{2}}d_{xy,\sigma _{1}} 
\nonumber \\
&+&J_{b}\sum_{\zeta }\sum_{\sigma _{1},\sigma _{2}}d_{3z^{2}-r^{2},\sigma
_{1}}^{\dagger }d_{\zeta ,\sigma _{2}}^{\dagger }d_{3z^{2}-r^{2},\sigma
_{2}}d_{\zeta ,\sigma _{1}}  \nonumber \\
&+&J_{e}\sum_{\chi }(d_{\chi ,\uparrow }^{\dagger }d_{\chi ,\downarrow
}^{\dagger }d_{3z^{2}-r^{2},\downarrow }d_{3z^{2}-r^{2},\uparrow }+\text{H.c.%
})  \nonumber \\
&+&J_{t}\sum_{\beta \neq \gamma }d_{\beta ,\uparrow }^{\dagger }d_{\beta
,\downarrow }^{\dagger }d_{\gamma ,\downarrow }d_{\gamma ,\uparrow }, 
\nonumber \\
&+&J_{t}\sum_{\zeta }(d_{x^{2}-y^{2},\uparrow }^{\dagger
}d_{x^{2}-y^{2},\downarrow }^{\dagger }d_{\zeta ,\downarrow }d_{\zeta
,\uparrow }+\text{H.c.})  \nonumber \\
&+&J_{a}(d_{x^{2}-y^{2},\uparrow }^{\dagger }d_{x^{2}-y^{2},\downarrow
}^{\dagger }d_{xy,\downarrow }d_{xy,\uparrow }+\text{H.c.})  \nonumber \\
&+&J_{b}\sum_{\zeta }(d_{3z^{2}-r^{2},\uparrow }^{\dagger
}d_{3z^{2}-r^{2},\downarrow }^{\dagger }d_{\zeta ,\downarrow }d_{\zeta
,\uparrow }+\text{H.c.})\text{ \ \ \ \ \ \ \ \ \ \ }  \nonumber \\
&+&\lambda \sum_{\sigma _{1},\sigma _{2}}[2(n_{yz,\sigma _{1}}-n_{zx,\sigma
_{1}})(d_{3z^{2}-r^{2},\sigma _{2}}^{\dagger }d_{x^{2}-y^{2},\sigma _{2}}+%
\text{H.c.})  \nonumber \\
&&-2(d_{3z^{2}-r^{2},\sigma _{1}}^{\dagger }d_{xy,\sigma _{1}}+\text{H.c.}%
)(d_{zx,\sigma _{2}}^{\dagger }d_{yz,\sigma _{2}}+\text{H.c.})  \nonumber \\
&&+\sqrt{3}(d_{x^{2}-y^{2},\sigma _{1}}^{\dagger }d_{zx,\sigma _{1}}+\text{%
H.c.})(d_{xy,\sigma _{2}}^{\dagger }d_{yz,\sigma _{2}}+\text{H.c.}) 
\nonumber \\
&&-\sqrt{3}(d_{x^{2}-y^{2},\sigma _{1}}^{\dagger }d_{yz,\sigma _{1}}+\text{%
H.c.})(d_{xy,\sigma _{2}}^{\dagger }d_{zz,\sigma _{2}}+\text{H.c.}) 
\nonumber \\
&&+(d_{3z^{2}-r^{2},\sigma _{1}}^{\dagger }d_{zx,\sigma _{1}}+\text{H.c.}%
)(d_{xy,\sigma _{2}}^{\dagger }d_{yz,\sigma _{2}}+\text{H.c.})  \nonumber \\
&&+(d_{3z^{2}-r^{2},\sigma _{1}}^{\dagger }d_{yz,\sigma _{1}}+\text{H.c.}%
)(d_{xy,\sigma _{2}}^{\dagger }d_{zx,\sigma _{2}}+\text{H.c.})  \nonumber \\
&&+(d_{3z^{2}-r^{2},\sigma _{1}}^{\dagger }d_{zx,\sigma _{2}}^{\dagger
}d_{x^{2}-y^{2},\sigma _{2}}d_{zx,\sigma _{1}}  \nonumber \\
&&-d_{3z^{2}-r^{2},\sigma _{1}}^{\dagger }d_{yz,\sigma _{2}}^{\dagger
}d_{x^{2}-y^{2},\sigma _{2}}d_{yz,\sigma _{1}}+\text{H.c.})]  \nonumber \\
&+&\lambda [ (d_{x^{2}-y^{2},\downarrow }d_{3z^{2}-r^{2},\uparrow
}-d_{x^{2}-y^{2},\uparrow }d_{3z^{2}-r^{2},\downarrow })  \nonumber \\
&&\times (d_{zx,\uparrow }^{\dagger }d_{zx,\downarrow }^{\dagger
}-d_{yz,\uparrow }^{\dagger }d_{yz,\downarrow }^{\dagger })+\text{H.c.}]
\label{hi}
\end{eqnarray}


\begin{thebibliography}{99}
\bibitem{note} We use the primes in $a_{1g}^{\prime }$ and $e_{g}^{\prime }$
to differentiate the irreducible representations of the point group $D_{3d}$
of the octahedron compressed along a $_3$ axis, to those of the point group 
$O_h$ of the regular octahedron.
\end{thebibliography}
\end{document}